\documentclass[pra,reprint,a4paper,superscriptaddress,citeautoscript,floatfix]{revtex4-1}

\pdfoutput=1
\usepackage[charter]{mathdesign}
\usepackage{amsmath}
\usepackage[pdftex]{graphicx}
\usepackage{microtype}
\usepackage{bm}

\usepackage[svgnames]{xcolor}
\usepackage[
	colorlinks=True,linkcolor=DarkRed,citecolor=ForestGreen,urlcolor=MediumBlue,
	pdf startview=FitH,bookmarks=False,pdfpagemode=UseNone
]{hyperref}

\begin{document}
\def\cutoff{\hbar\omega_{\mathrm{D}}}

\title{Periodicity of superconducting shape resonances in thin films}

\author{D. Valentinis}
\affiliation{Institute for Theoretical Condensed Matter physics, Karlsruhe Institute of Technology, Wolfgang-Gaede Stra{\ss}e 1, 76131 Karlsruhe, Germany}
\author{C. Berthod}
\affiliation{Department of Quantum Matter Physics, University of Geneva, 24 quai Ernest-Ansermet, 1211 Geneva, Switzerland}

\date{August 25, 2020}

\begin{abstract}

The pairing temperature of superconducting thin films is expected to display, within the Bardeen--Cooper--Schrieffer theory, oscillations as a function of the film thickness. We show that the pattern of these oscillations switches between two different periodicities at a density-dependent value of the superconducting coupling. The transition is most abrupt in the anti-adiabatic regime, where the Fermi energy is less than the Debye energy. To support our numerical data, we provide new analytical expressions for the chemical potential and the pairing temperature as a function of thickness, which only differ from the exact solution at weak coupling by exponentially-small corrections.

\end{abstract}

\maketitle

\section{Introduction}

Since the pioneering study of Thompson and Blatt raised hopes to observe improved critical temperature in thin films made of superconducting materials \cite{Blatt-1963, *Thompson-1963}, a large number of experimental \cite{Strongin-1965, *Strongin-1968, Abeles-1966, Alekseevskii-1971, Komnik-1970, Orr-1984, Bao-2005, Eom-2006, Qin-2009, Kang-2011, Kim-2012, Navarro-2016, Bose-2010, Pasztor-2017, Pinto-2018, Yang-2018, Uchihashi-2016} and theoretical \cite{Kirzhnits-1965, Kresin-1966, Shapoval-1967, Perali-1996, *Cariglia-2016, Hwang-2000, Croitoru-2007, *Shanenko-2007, *Shanenko-2008, Araujo-2011, Chen-2012a, Chen-2012b, Mayoh-2014, Bianconi-2014, Romero-Bermudez-2014a, Romero-Bermudez-2014b, Bekaert-2017, Virtanen-2019} works have followed up on this idea. Thanks to the quantum confinement along one direction, the thin-film geometry splits the three-dimensional dispersion law of the superconductor into a set of two-dimensional subbands. The energy separation between the subbands varies with changing film thickness such that the Fermi level, which is fixed by the bulk electron density, must adjust as well. In the Thompson--Blatt model (a free-electron like metal confined in the film by hard walls), the critical temperature varies with reducing film thickness, drawing a sawtooth-like increase (Fig.~\ref{fig:fig1}), where jumps occur each time the Fermi level crosses the bottom of a subband. These quantum oscillations have become known as superconducting shape resonances. The resulting ``period'' (actually a wavelength) of critical-temperature oscillations is
	\begin{equation}\label{eq:Lambda0}
		\Lambda_0=\frac{\pi}{k_{\mathrm{F}}}\approx n^{-1/3},
	\end{equation}
where $k_{\mathrm{F}}$ and $n$ are the bulk Fermi wave vector and electron density, respectively. For typical metallic densities of order $10^{22}$~cm$^{-3}$, the expected oscillations period is a few Angstr\"{o}m. The period $\Lambda_0$ obtained by Thompson and Blatt tracks \emph{discontinuities} of the critical temperature $T_c$ versus film thickness $L$. These discontinuities arise due to a simplification adopted when solving the Bardeen--Cooper--Schrieffer (BCS) gap equation, while the exact dependence $T_c(L)$ is \emph{continuous} \cite{Valentinis-2016-2}. The simplification consists in ignoring that, when the Fermi energy is sufficiently close to the bottom of a subband, the frequency-dependent pairing interaction is cut by the subband edge rather than by the ordinary Debye cutoff $\cutoff$. Although the exact function $T_c(L)$ is continuous, its first derivative $dT_c/dL$ has discontinuities when the bottom of a subband coincides with the upper edge of the interaction window, i.e., rather than triggering a discontinuity of $T_c$ when it crosses the subband edge, the Fermi level triggers a discontinuity of $dT_c/dL$ when it reaches $\cutoff$ \emph{below} the subband edge. This leads to a corrected period \cite{Valentinis-2016-2},
	\begin{equation}\label{eq:Lambda}
		\Lambda=\frac{\pi}{\sqrt{k_{\mathrm{F}}^2+2m\omega_{\mathrm{D}}/\hbar}}
		\propto\frac{1}{\sqrt{E_{\mathrm{F}}+\cutoff}},
	\end{equation}
which tracks the discontinuities of $dT_c/dL$. The exact period (\ref{eq:Lambda}) is shorter than the Thompson--Blatt result (\ref{eq:Lambda0}), although both coincide in the adiabatic limit $E_{\mathrm{F}}\gg\cutoff$. Equations~(\ref{eq:Lambda0}) and (\ref{eq:Lambda}) are asymptotic results obtained in the weak-coupling regime $\lambda\ll1$, where $\lambda$ is the dimensionless coupling constant for pairing. In this limit, $T_c$ approaches zero and the chemical potential at $T_c$ is close to the zero-temperature Fermi energy. Furthermore, these expressions are valid for large $L$, where the period becomes well defined and the Fermi energy approaches the bulk value.

Simulations performed at intermediate to strong coupling show that Eq.~(\ref{eq:Lambda}) works in this regime as well \footnote{Small deviations occur at strong coupling due to the departure of the chemical potential from the noninteracting value, ignored in deriving Eq.~(\ref{eq:Lambda}); see Ref.~\onlinecite{Valentinis-2016-2}.}. The discontinuities of $dT_c/dL$ are large in that case (in a sense to be made precise below) and the $T_c(L)$ curve has cusps pointing downward at the discontinuities, separated by maxima in between each cusp (Fig.~\ref{fig:fig2}). Since the optimal condition to observe the difference between Eqs.~(\ref{eq:Lambda0}) and (\ref{eq:Lambda}) is the anti-adiabatic regime $E_{\mathrm{F}}\lesssim\cutoff$, which is often associated with strong coupling \cite{Pietronero-1992, Banacky-2009, Sadovskii-2019}, it is interesting that Eq.~(\ref{eq:Lambda}) is valid beyond weak coupling. Of course, the applicability of the static BCS approach is not guaranteed for these cases. Luckily, there exists low-density systems such as $n$-doped SrTiO$_3$ which, albeit falling into the class of anti-adiabatic superconductors \cite{Gorkov-2016}, have low values of the coupling constants \cite{vanderMarel-2011, Swartz-2018, Thiemann-2018, Valentinis-2017, Gastiasoro-2020}. Simulations of the $T_c(L)$ curves performed at low values of $\lambda$ show, however, that the oscillation pattern changes as $\lambda\to0$. The size of the discontinuities in $dT_c/dL$ decreases and the relative amplitude of the oscillations in $T_c(L)$ increases. While the separation between discontinuities continues to be described by Eq.~(\ref{eq:Lambda}), the new oscillation pattern is not controlled by these discontinuities any more and approaches a period given, somewhat surprisingly, by Eq.~(\ref{eq:Lambda0}). Thus, in the anti-adiabatic regime, where Eq.~(\ref{eq:Lambda}) would suggest that the period of $T_c$ oscillations becomes independent of the density, this is true only for moderate to strong coupling, while the density dependence given by Eq.~(\ref{eq:Lambda0}) reappears at weak coupling. This is the main message of the present paper, which we elaborate in the following.

\section{Model and results}

We consider a simple BCS superconductor with parabolic dispersion and a local electron-electron attraction, that is confined by two parallel hard walls. The more realistic case of a finite-depth potential well can be treated similarly at the cost of introducing one additional parameter, but this plays a marginal role in the question of the periodicity discussed here. The value of the critical temperature $T_c$ is found by solving the following set of coupled equations:
	\begin{subequations}\label{eq:BCS}\begin{align}
		\label{eq:BCS1}
		n&=\frac{mk_{\mathrm{B}}T_c}{\pi\hbar^2L}\sum_q\ln\left(1+e^{\frac{\mu-E_{q}}
		{k_{\mathrm{B}}T_c}}\right),\\
		\label{eq:BCS2}
		\Delta_p&=\sum_qV_{pq}\Delta_q\frac{m}{2\pi\hbar^2}
		\int_{-\cutoff}^{\cutoff}dE\,\theta(\mu+E-E_q)\,
		\frac{\tanh\left(\frac{E}{2k_{\text{B}}T_c}\right)}{2E}.
	\end{align}\end{subequations}
These equations may be derived from the most general Gor'kov mean-field expressions by linearizing them at $T_c$, where all order parameters vanish, and specializing to a separable BCS-like pairing interaction (see Appendix of Ref.~\onlinecite{Valentinis-2016-2}). Equation~(\ref{eq:BCS1}) sets the chemical potential $\mu(n,L,T_c)$, such as to keep the electron density fixed when $L$ and $T_c$ vary. The $q$ sum runs over all nonzero positive integers, with $E_q=\frac{\hbar^2}{2m}\left(\frac{q\pi}{L}\right)^2$ giving the minima of the subbands in the quantum well. The simple form of the density equation with a logarithm results after summing the Fermi occupation factors for the momenta parallel to the confinement walls. Equation~(\ref{eq:BCS2}) is the linearized gap equation at $T_c$, where the pairing order parameters $\Delta_q$ in all subbands vanish. The 3D electron-electron attraction has the same matrix element $V$ between all states having energy within the range $[-\cutoff,+\cutoff]$ from the chemical potential. Equation~(\ref{eq:BCS2}) is, however, written in the basis of the quantum-well eigenstates, where the matrix elements are no longer all identical, but are larger for the intra-subband processes than for the inter-subband ones: $V_{pq}=\frac{V}{L}(1+\delta_{pq}/2)$ \cite{Blatt-1963, *Thompson-1963, Valentinis-2016-2}. The integration variable $E$ spans the dynamical range of the interaction and accounts for the energy gained by pairing states of subband $q$ in that range, weighted by $m/(2\pi\hbar^2)$, which is the density of states of the subband. When $\mu+E<E_q$, the energy $E$ falls below the subband, where there are no states to pair, hence the Heaviside function for removing that energy window from the integral.

The model has five parameters ($m$, $V$, $\cutoff$, $n$, $L$), which can be reduced to four by using $\cutoff$ as the unit of energy. Following Ref.~\onlinecite{Valentinis-2016-1}, we define a dimensionless density parameter:
	\begin{equation}\label{eq:ntilde}
		\tilde{n}=\frac{n}{2[m\omega_{\mathrm{D}}/(2\pi\hbar)]^{3/2}}
		=\frac{4}{3\sqrt{\pi}}\left(\frac{E_{\mathrm{F}}}{\cutoff}\right)^{3/2}.
	\end{equation}
It is seen that $\tilde{n}$ is not, strictly speaking, a measure of the density---for instance, at fixed physical density, $\tilde{n}$ changes if the mass of the particles changes---but rather a measure of the adiabatic ratio $E_{\mathrm{F}}/\cutoff$. The value $\tilde{n}\approx 0.75$ marks the transition between the anti-adiabatic regime $E_{\mathrm{F}}<\cutoff$ and the adiabatic regime $E_{\mathrm{F}}>\cutoff$. The dimensionless pairing strength is usually measured by the product of the interaction with the 3D density of states at the chemical potential, $\lambda=VN(\mu)$. This definition is impractical when $\mu$ is adjusted self-consistently and Ref.~\onlinecite{Valentinis-2016-1} used instead $\bar{\lambda}=VN(\cutoff)$. With the latter convention, the values of the coupling constant are not easily compared with experimentally determined values. In the present paper, we use the more conventional definition $\lambda=VN(E_{\mathrm{F}})$, where $E_{\mathrm{F}}$ is computed from $n$ using noninteracting-electron expressions, like in Eq.~(\ref{eq:ntilde}). In terms of the model parameters, the coupling constant is
	\begin{equation}\label{eq:lambda}
		\lambda=\frac{mV}{2\pi\hbar^2}\left(\frac{3n}{\pi}\right)^{1/3}.
	\end{equation}
With the definitions Eqs.~(\ref{eq:ntilde}) and (\ref{eq:lambda}), the coupled Eqs.~(\ref{eq:BCS}) only involve the four parameters $m$, $\lambda$, $\tilde{n}$, and $L$.

\begin{figure}[tb]
\includegraphics[width=\columnwidth]{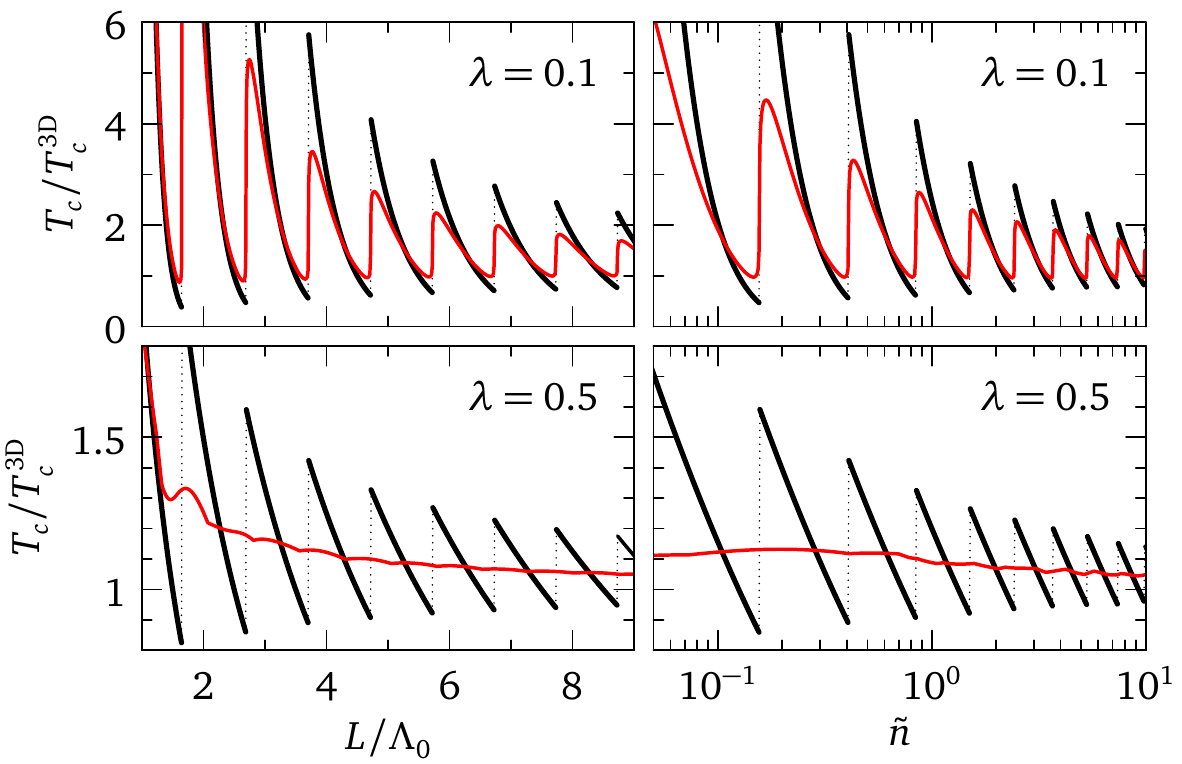}
\caption{\label{fig:fig1}
Variations of the BCS critical temperature relative to the 3D bulk value showing shape resonances versus film thickness at fixed electron density (left panels, $\tilde{n}=1$) and versus electron density at fixed film thickness [right panels, $L=5\Lambda_0(\tilde{n}=1)$]. The black curves show the Thompson--Blatt result [Eqs.~(\ref{eq:TB}) or (\ref{eq:TBclosed})] while the red curves show the exact result [Eqs.~(\ref{eq:BCS})]. The exact curves approach the Thompson--Blatt curves at weak coupling (upper panels).
}
\end{figure}

Two simplifications are sometimes made to Eqs.~(\ref{eq:BCS}): The density equation is replaced by its zero-temperature limit and in Eq.~(\ref{eq:BCS2}), $\theta(\mu+E-E_q)$ is replaced by $\theta(\mu-E_q)$. The resulting simplified equations are:
	\begin{subequations}\label{eq:TB}\begin{align}
		\label{eq:TB1}
		n&=\frac{m}{\pi\hbar^2L}\sum_q\max\left(0,\mu-E_q\right)\\
		\label{eq:TB2}
		\Delta_p&=\sum_qV_{pq}\Delta_q
		\frac{m}{2\pi\hbar^2}\theta(\mu-E_q)\int_{-\cutoff}^{\cutoff}dE\,
		\frac{\tanh\left(\frac{E}{2k_{\text{B}}T_c}\right)}{2E}.
	\end{align}\end{subequations}
By solving Eqs.~(\ref{eq:TB}) numerically, we obtain the discontinuous variations of $T_c$ shown in Fig.~\ref{fig:fig1} as black lines. This is reminiscent of the Thompson--Blatt results who, rather than solving Eqs.~(\ref{eq:TB}) at $T_c$, computed the order parameters at $T=0$ using equivalent simplifications. The system of Eqs.~(\ref{eq:TB}) admits a closed solution that reproduces accurately the data shown in the figure (see Appendix~\ref{app:TB}). Figure~\ref{fig:fig1} also shows the solution of Eqs.~(\ref{eq:BCS}) in red for comparison. There are significant differences, but the red lines seem to approach the approximate result at weak coupling.

\begin{figure}[tb]
\includegraphics[width=\columnwidth]{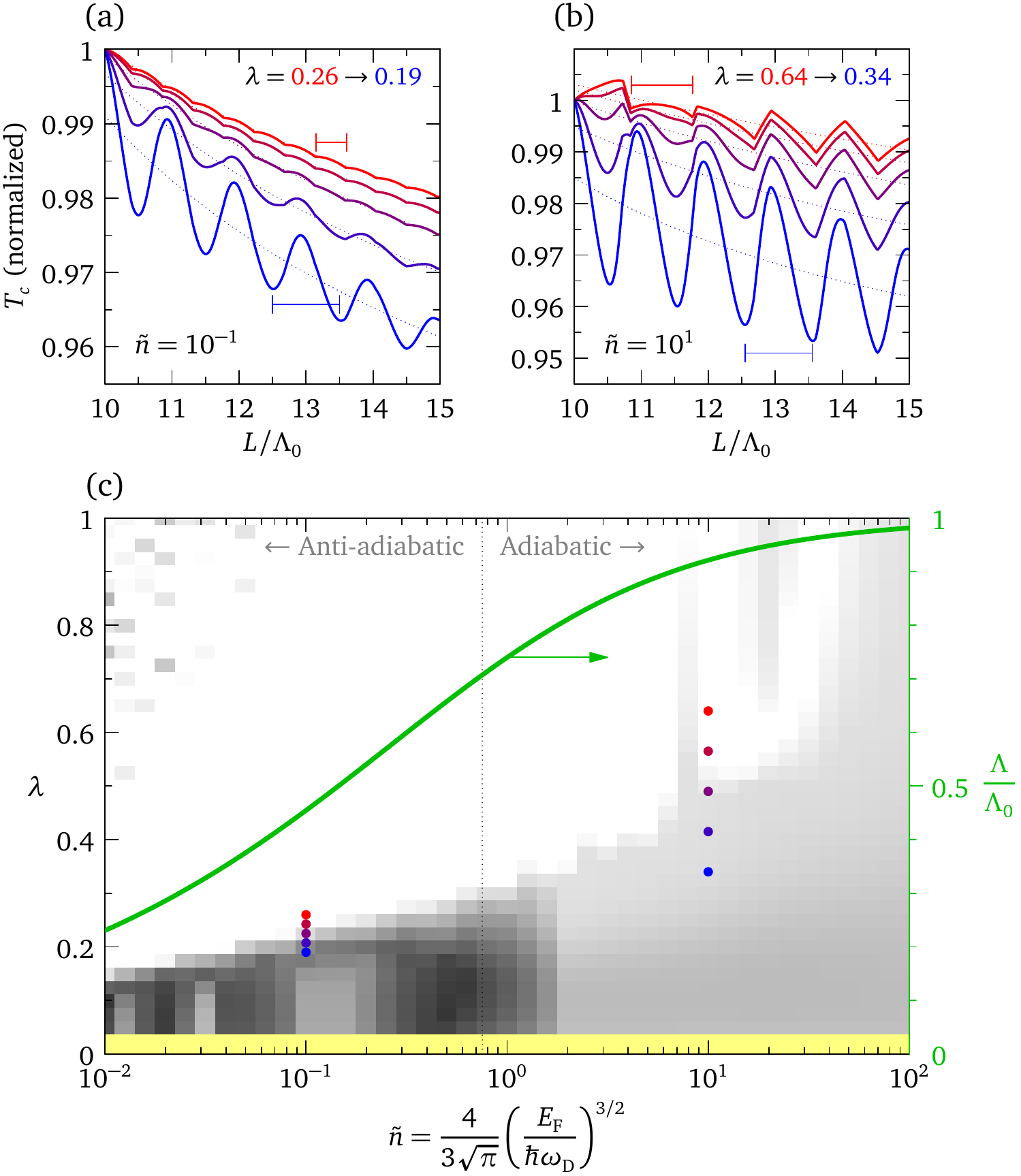}
\caption{\label{fig:fig2}
(a), (b) Evolution of $T_c$ with film thickness in the anti-adiabatic (a) and adiabatic (b) regimes. $T_c(L)$ is normalized to its value at $L=10\Lambda_0$. Different curves correspond to different coupling constants, as indicated by the dots in (c). The horizontal red and blue bars show $\Lambda$ and $\Lambda_0$, respectively. The dotted lines show the fitted background. (c) Illustration of the crossover from Eq.~(\ref{eq:Lambda}) (bright) to Eq.~(\ref{eq:Lambda0}) (dark) with decreasing $\lambda$ across the anti-adiabatic and adiabatic regimes. The gray scale shows the ratio of Fourier components at $2\pi/\Lambda$ and $2\pi/\Lambda_0$ (see text). $T_c$ is smaller than machine precision in the yellow region. The green curve (right scale) shows that $\Lambda$ and $\Lambda_0$ become difficult to distinguish in the adiabatic regime. All calculations are done for a mass equal to the bare electron mass.
}
\end{figure}

Figures~\ref{fig:fig2}(a) and \ref{fig:fig2}(b) show some more results from Eqs.~(\ref{eq:BCS}), with $T_c(L)$ displaying quantum oscillations on top of a background that increases with decreasing $L$. At sufficiently large coupling (red curves), the oscillation period is set by the discontinuities of $dT_c/dL$, which correspond to downward-pointing cusps, leading to Eq.~(\ref{eq:Lambda}). In the adiabatic regime [Fig.~\ref{fig:fig2}(b)], additional discontinuities appear in between, that occur when the Fermi level is $\cutoff$ \emph{above} the bottom of a subband \cite{Valentinis-2016-2}. As the coupling is reduced, the discontinuities of $dT_c/dL$ are suppressed and the quantum oscillations display the period $\Lambda_0$ (blue curves). To measure the evolution of the period as a function of coupling, we calculate the dependence $T_c(L)$ for $10\Lambda_0<L<100\Lambda_0$, we remove the background by fitting it to the form $T_c(\infty)+1/(a+bL^c)$, and we compute the cosine transform of the remaining function. The ratio of the Fourier coefficients at $2\pi/\Lambda$ and $2\pi/\Lambda_0$ indicates the dominant period. Repeating this calculation at each density and coupling, we obtain the data shown in Fig.~\ref{fig:fig2}(c). Although this measure is somewhat noisy, it shows well the transition from the period Eq.~(\ref{eq:Lambda}) to the period Eq.~(\ref{eq:Lambda0}) as the coupling is reduced. The transition is sharp in the anti-adiabatic regime and becomes more and more gradual as one enters the adiabatic regime. At large $\tilde{n}$, both periods become similar and their difference reaches the resolution limit of our Fourier transform.

\begin{figure}[b]
\includegraphics[width=\columnwidth]{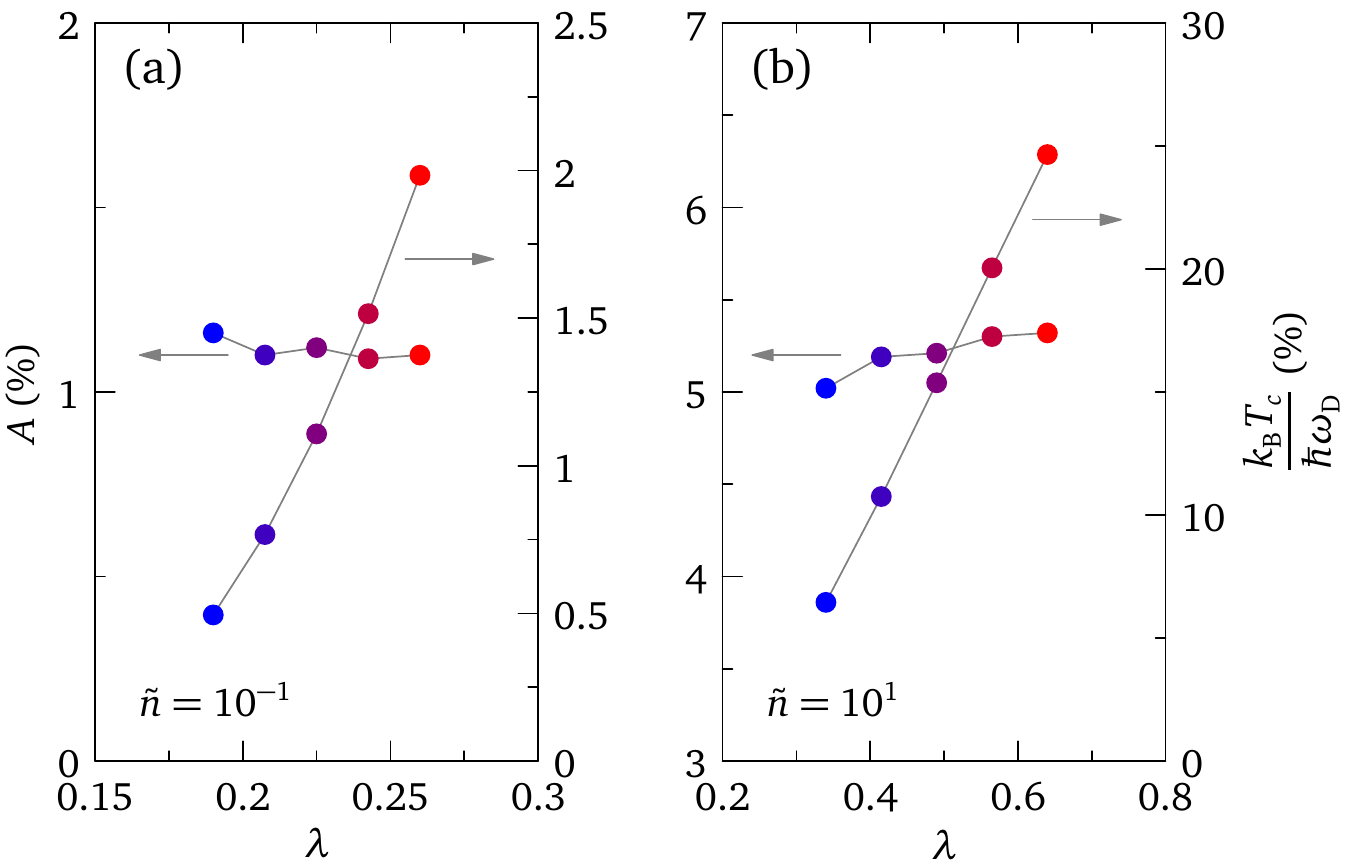}
\caption{\label{fig:fig3}
Evolution of the discontinuity measure (left scales) and critical temperature at the discontinuity (right scales) across the transition from $\Lambda$ to $\Lambda_0$ periodicity in the anti-adiabatic (a) and adiabatic (b) regimes.
}
\end{figure}

The change of period is associated with a suppression of the discontinuities in $dT_c/dL$. To quantify the strength of the discontinuities, we consider the dimensionless quantity,
	\begin{equation}\label{eq:A}
		A=\frac{(dT_c/dL)_+-(dT_c/dL)_-}{T_c/L},
	\end{equation}
which can be evaluated at each discontinuity of $dT_c/dL$. Figure~\ref{fig:fig3} shows this quantity calculated with the data plotted in Figs.~\ref{fig:fig2}(a) and \ref{fig:fig2}(b) at the first discontinuity following $L=10\Lambda_0$. It is seen that $A$ is approximately constant across the transition between the two periods. This means that the size of the discontinuity scales like $T_c$ and therefore drops exponentially at weak coupling. The evolution of $T_c$ is also shown in Fig.~\ref{fig:fig3} for comparison.

\begin{figure}[tb]
\includegraphics[width=\columnwidth]{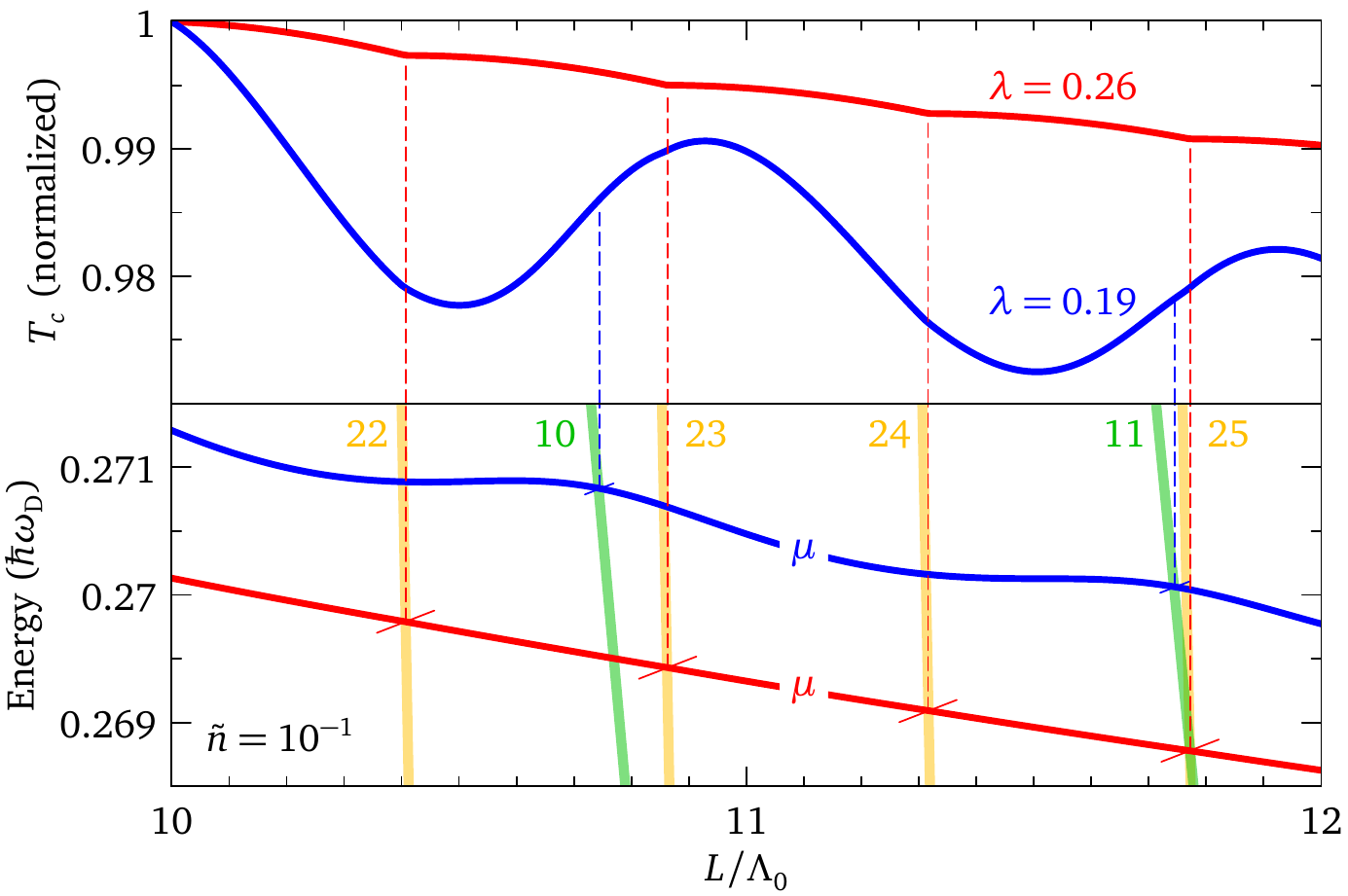}
\caption{\label{fig:fig4}
Critical temperature as in Fig.~\ref{fig:fig2}(a) for two values of $\lambda$ (top) and corresponding evolution of the chemical potential (bottom). The thick green lines show the minima of subbands 10 and 11 as they vary with $L$ and the lines labeled 22 to 25 indicate an energy lying $\cutoff$ below the corresponding subbands.
}
\end{figure}

When the discontinuities become subdominant on the $T_c(L)$ curve and the periodicity turns to Eq.~(\ref{eq:Lambda0}), it is tempting to attribute each $T_c$ maximum to a coincidence between the chemical potential and the edge of a subband. This is not the case, as Fig.~\ref{fig:fig4} shows for the data of Fig.~\ref{fig:fig2}(a). Before presenting this figure, we point out that self-consistency of the chemical potential is crucial: the thicknesses at which $\mu(L)$ presents a discontinuous derivative are different at $T=0$ and $T=T_c$ \cite{Valentinis-2016-2}. Therefore, the conclusions drawn from analyzing the shape resonances of the excitation gap at $T=0$ \cite{Croitoru-2007, *Shanenko-2007, *Shanenko-2008} may differ from those drawn from the $T_c$ curve. Figure~\ref{fig:fig4} and all our numerical calculations involve the chemical potential calculated self-consistently at $T_c$. To describe this figure, we start at $L/\Lambda_0=12$ with $\lambda=0.26$ (red curves). The chemical potential lies inside the 11th subband. Upon reducing $L$, everything else held fixed, the electron density would increase like $1/L$ due to compression, such that a lowering of the chemical potential would be needed to compensate. However, all subbands move up in energy like $1/L^2$ with reducing thickness: the ensuing loss of states overweights the compression such that the chemical potential must follow the trend of the bands and increase like $1/L$. The critical temperature also has an increasing trend because the pairing matrix elements vary like $1/L$ \cite{Blatt-1963, *Thompson-1963}. Below $L/\Lambda_0=11.8$, the 25th subband at energy $\mu+\cutoff$ ceases contributing to pairing and this induces a cusp in $T_c$ and the discontinuity in $dT_c/dL$. Accidentally, this is also the point where the chemical potential leaves the 11th subband, but this crossing imprints no signature in $T_c$, as can be seen when $\mu$ crosses the 10th subband at lower thickness. For $\lambda=0.19$ (blue curves), the critical temperature is lower and the chemical potential is correspondingly higher. For the rest, a precise interpretation seems difficult. Starting from $L/\Lambda_0=12$, both $T_c$ and $\mu$ show an increasing trend like for stronger coupling. However, near $L/\Lambda_0=11.9$, $T_c$ starts to decrease before the chemical potential leaves the 11th subband and then goes through a minimum at a thickness where $\mu$ has no obvious coincidence with the subband energies. The feature in $T_c(L)$ which seems to correlate best with $\mu$ crossing a subband is a zero of the second derivative, where the curvature changes from negative to positive with decreasing $L$. The same conclusion is reached in the adiabatic regime with the data of Fig.~\ref{fig:fig2}(b). 

Figure~\ref{fig:fig1} suggests that the exact $T_c$ at weak coupling interpolates smoothly across the discontinuities of the approximate result. These discontinuities occur when $\mu_0$ crosses a subband edge, where $\mu_0$ is the chemical potential given by Eq.~(\ref{eq:TB1}). Provided that the difference between the exact $\mu$ and $\mu_0$ becomes negligible at weak coupling, this would explain the coincidence between the curvature changes of $T_c(L)$ and $\mu$ crossing a subband edge. In Appendix~\ref{app:weak}, we show that the exact chemical potential from Eqs.~(\ref{eq:BCS}) indeed approaches the value $\mu_0$ given by Eq.~(\ref{eq:TB1}) when $T_c\to0$, unless the vanishing of $T_c$ is driven by taking another limit, either $L\to0$ or $n\to0$. In the latter cases, $\mu(T_c=0)\neq\mu_0$ \cite{Valentinis-2016-1, Valentinis-2016-2}. But for any finite $L$ and $n$, we find that the deviation of $\mu(T_c\to0)$ from $\mu_0$ is exponentially small in $k_{\mathrm{B}}T_c/\mu_0$. Furthermore, we also show, based on a closed solution, that the $T_c$ resulting from Eqs.~(\ref{eq:BCS}) approaches the one from Eqs.~(\ref{eq:TB}) with corrections that are exponentially small for $\lambda\to0$ (except in the two limits mentioned above). This allows us to conclude that in the regime where the solution of Eqs.~(\ref{eq:BCS}) oscillates with the period $\Lambda_0$, the inflection points where the curvature changes from positive to negative with increasing $L$ signal the population of a new subband.

\begin{figure}[tb]
\includegraphics[width=\columnwidth]{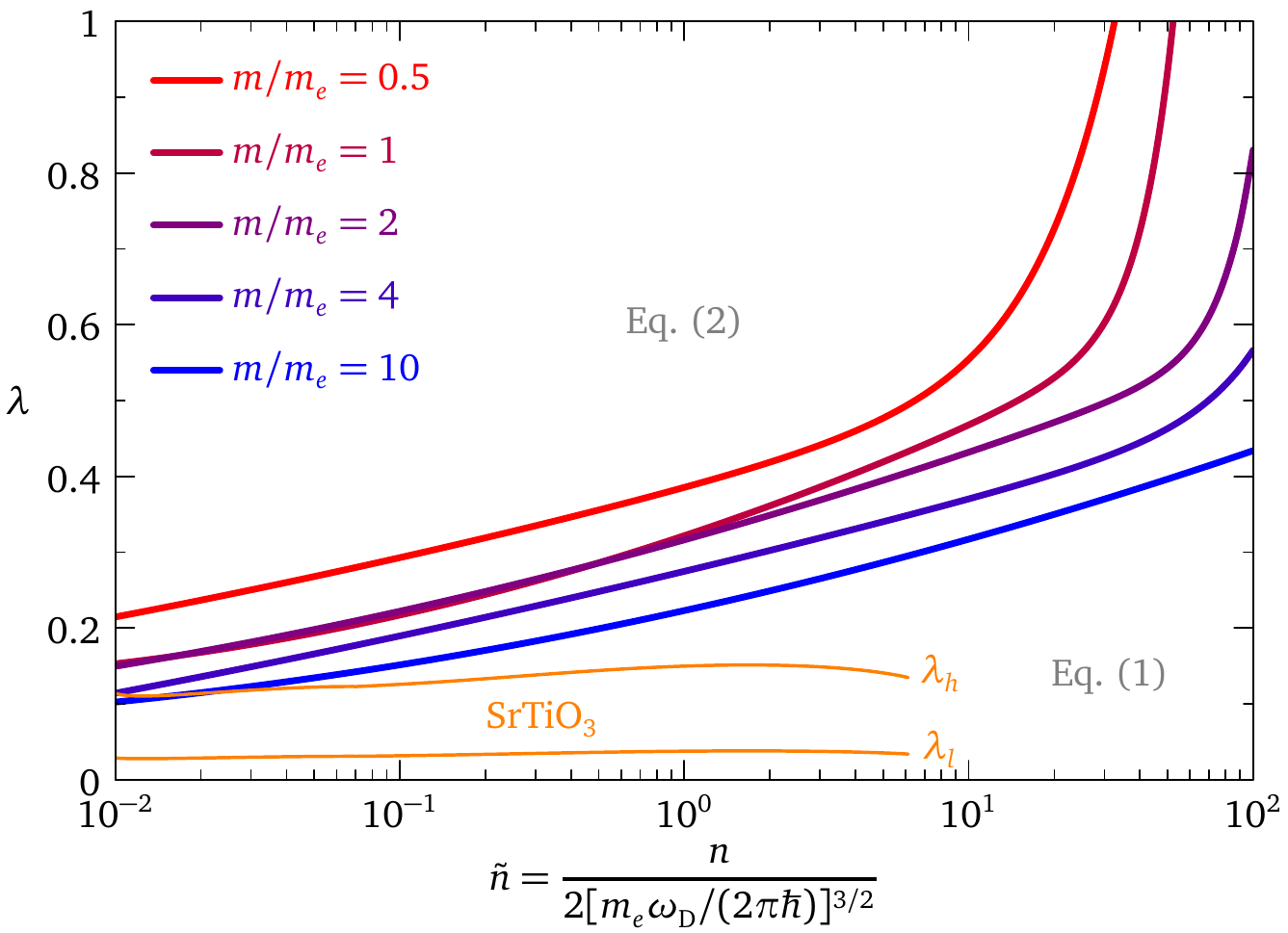}
\caption{\label{fig:fig5}
Mass dependence of the boundary between the periodicities given by Eqs.~(\ref{eq:Lambda0}) and (\ref{eq:Lambda}). The orange lines show the coupling constants in the light and heavy bands of SrTiO$_3$, as determined in Ref.~\onlinecite{Valentinis-2017}.
}
\end{figure}

The boundary between the two periodicities in Fig.~\ref{fig:fig2}(c) depends on the carrier mass. In Fig.~\ref{fig:fig5}, we show the boundary extracted from Fig.~\ref{fig:fig2}(c), together with boundaries obtained with other values of the mass. To compare different masses, we normalize the density on the horizontal axis using the bare electron mass $m_e$ in all cases. As the mass increases, the domain of Thompson--Blatt periodicity shrinks and moves to higher densities. We also show in Fig.~\ref{fig:fig5} the density-dependent coupling constants $\lambda_l$ and $\lambda_h$ for SrTiO$_3$, as determined in Ref.~\onlinecite{Valentinis-2017} for the light ($l$) and heavy ($h$) bands with masses $m_l=m_e$ and $m_h=4m_e$, respectively. As it turns out, in the whole range of densities, SrTiO$_3$ falls in the regime of the Thompson--Blatt periodicity Eq.~(\ref{eq:Lambda0}). Therefore, in spite of the fact that this low-density material lies well within the anti-adiabatic regime, thin films of doped SrTiO$_3$ are expected to display oscillations of $T_c$ with the period Eq.~(\ref{eq:Lambda0}), because of the low coupling \cite{Kim-2012, Valentinis-2018}. We discuss the case of SrTiO$_3$ further below.

\section{Discussion and conclusion}

In summary, we have studied a model of BCS superconductor confined to a thin film and we report a change from a short to a long oscillation period in the pattern of critical-temperature shape resonances. The cross-over from one period to the other occurs at a density- and mass-dependent value of the coupling strength. The long period is found at weaker coupling, larger carrier density, and lower carrier mass. The short period tracks discontinuities in the derivative of the variation of $T_c$ versus film thickness. These discontinuities vanish exponentially as the coupling is reduced and the long periodicity emerges, which only depends on the carrier density. We support our numerical findings with new analytical results.

The experimental demonstration of shape resonances in thin films requires observing oscillations of $T_c$ with varying the film thickness. For simple band structures, the oscillation pattern is linked with microscopic parameters of the bulk material, allowing one to check that the variations of $T_c$ are indeed controlled by the quantum confinement. A clear-cut demonstration of this effect in superconducting thin films has proven difficult. While a mere increase or decrease of $T_c$ with changing thickness is routinely observed, this is not, per se, proof that confinement effects of the kind discussed here do occur. These variations can be attributed to other causes like proximity effects \cite{Pinto-2018} or the tuning of an alternate order competing with superconductivity \cite{Yang-2018}.

Shape resonances require electronic coherence over the film thickness. This is manifested in Eqs.~(\ref{eq:BCS}) by the subband quantization ($E_q$) and pairing matrix elements ($V_{pq}$), that both require the electronic wave functions to coherently feel the two boundaries of the film. Depending on the ratio between the inelastic (momentum-relaxing) electron mean-free path $\lambda$ and the film thickness $L$, different mesoscopic transport regimes are realized and distinct measures of electronic coherence are relevant \cite{[{See, e.g., }] Ferry-1997}. In the clean regime $\lambda>L$, attainable at low temperatures in systems with a low concentration of defects, the electronic response is coherent over the entire film thickness and $\lambda$ itself provides the electronic coherence scale. On the other hand, in the dirty regime $\lambda<L$, phase coherence can still be preserved over the film if the phase-coherence length $\ell_{\phi}>L$. The condition $\ell_{\phi}>L$ can be rearranged using Eq.~(\ref{eq:Lambda0}) as $L/\Lambda_0<2\ell_{\phi}/\lambda_{\mathrm{F}}$, where $\lambda_{\mathrm{F}}$ is the Fermi wavelength. The left-hand side $L/\Lambda_0$ can be interpreted as the number of resonances that can develop with increasing $L$, before $L$ exceeds $\ell_{\phi}$. Thus, as a rule of thumb, the number of observable resonances is expected to be twice the ratio of the phase-coherence length to the Fermi wavelength. In practice, the film thickness can only be varied by integer multiples of the lattice parameter $a$. For high-density metals with $k_{\mathrm{F}}\sim\pi/a$, the period $\Lambda_0\sim a$ is too short to be observed. The shape resonances should rather be searched in low-density metals with $\Lambda_0\gg a$. Low-density metals may lie in the anti-adiabatic regime, where $\Lambda_0$ is replaced by a shorter period $\Lambda$ given by Eq.~(\ref{eq:Lambda}) if the coupling is sufficiently strong. But the simulations show that the relative amplitude of $T_c$ oscillations is largest at weak coupling. Therefore, the optimal conjunction for an observation of shape resonances is a low-density metal with a weak superconducting coupling and a long electronic coherence length relative to the Fermi wavelength.

Elemental bismuth is the lowest-density superconductor with $n=3\times 10^{17}$~cm$^{-3}$ \cite{Prakash-2017} and probably the first metal in which quantum-confinement effects have been observed in the transport properties \cite{Ogrin-1966}, thanks to a very long mean free-path in the micrometer range. Due to a tiny carrier mass of order $10^{-3}m_e$, the Fermi energy is as large as 25~meV, to be compared with a Debye energy of 12~meV. These figures locate bismuth at $\tilde{n}\approx 2.3$, in the adiabatic side of Fig.~\ref{fig:fig2}, although this material is usually labeled as anti-adiabatic \cite{Prakash-2017, Koley-2017}. The expected period $\Lambda_0\sim150$~\AA{} is large but, unfortunately, in this material the level quantization opens a gap and destroys the metallic state for films thinner than 300~\AA{} \cite{Hoffman-1993}.

Another low-density superconductor is doped SrTiO$_3$ (STO), the first discovered oxide superconductor \cite{Schooley-1964, *Schooley-1965, *Koonce-1967}. Oxygen reduction and Nb doping allow one to vary the carrier density in a broad range covering three decades, from $3.5\times 10^{17}$ to $3.5\times 10^{20}$~cm$^{-3}$ \cite{Lin-2014, Collignon-2019}. Unlike in bismuth, the carrier mass is of the order of the bare electron mass, resulting in a range of $\tilde{n}$ values spanning the whole anti-adiabatic to adiabatic crossover from $\tilde{n}=0.01$ to $\tilde{n}=6$, as seen in Fig.~\ref{fig:fig5}. The figure also shows that the coupling constants are small. Hence, STO fulfills the conditions of being a low-density metal with a weak superconducting coupling. The expected $T_c$ oscillation period $\Lambda_0$ varies from $\sim 14$~nm at the lowest densities to $\sim 1.5$~nm at the highest ones. The transport mean-free path of STO single crystals reaches values above 200~nm at low $T$ \cite{Lin-2017}. For Nb-doped thin films in the dirty regime grown by pulsed laser deposition, the temperature dependence of the upper critical field indicates slightly lower values of $\ell_{\phi}$ in the range 70--130~nm \cite{Leitner-2000}. Similar figures were obtained for two-dimensional electron gases. A study of the universal conductance fluctuations in a surface electron gas made by ion-liquid gating undoped STO reports values of the phase-coherence length above 200~nm at $\sim0.4$~K. The magnetoresistance at the SrTiO$_3$/LaAlO$_3$ interface points to $\ell_{\phi}=157$~nm at 1.3~K \cite{Rakhmilevitch-2010}, indicating a good coherence of the electrons, in line with the recent observation of tunable confinement effects in the normal state \cite{Caputo-2020}. Since the interface electron gas displays a superconductivity similar to that of bulk STO \cite{Valentinis-2017}, it is not surprising that the typical electronic coherence lengths are also similar.

In conclusion, with $\ell_{\phi}\gg\Lambda_0\gg a$, doped SrTiO$_3$ stands out as a candidate of choice for the observation of superconducting shape resonances. At a typical density of $10^{20}$~cm$^{-3}$ with two bands occupied, the Fermi wavelength is of the order of 5~nm and the period $\Lambda_0\approx 2.5$~nm is six times longer that the lattice spacing. The conservative estimate $\ell_{\phi}\gtrsim50$~nm would then imply that up to 20 resonances may possibly be observed in thin films, by progressively reducing their thickness below $\sim 100$ unit cells  \cite{Valentinis-2018}.

\begin{acknowledgments}

We thank A. Bianconi, H. Boschker, M. Doria, S. Gariglio, B. Keimer, T. Loew, J. Mannhart, A. Perali, N. Poccia, J.-M. Triscone, D. van der Marel, and J. Zaanen for discussions. This work was supported by the Swiss National Science Foundation under Division II and the Early Postdoc Mobility Grant No. P2GEP2\textunderscore181450 (D. V.).

\end{acknowledgments}

\appendix

\section{Shape resonances in the Thompson--Blatt model}
\label{app:TB}

The enhancement of $T_c$ relative to the bulk value shown in Fig.~\ref{fig:fig1} with the black lines was computed by solving numerically Eqs.~(\ref{eq:TB}). These equations can also be solved (almost) exactly. We give here a closed formula that produces curves undistinguishable from the numerical data shown in Fig.~\ref{fig:fig1}. The integral on the right-hand side of Eq.~(\ref{eq:TB2}) is independent of the band index $q$ and can be evaluated using
	\begin{equation}\label{eq:A1}
		\int_{-\cutoff}^{\cutoff}dE\,\frac{\tanh\left(\frac{E}{2k_{\text{B}}T_c}\right)}{2E}\approx
		\ln\left(\frac{2e^{\gamma}}{\pi}\frac{\cutoff}{k_{\mathrm{B}}T_c}\right).
	\end{equation}
The relation becomes exact only in the limit $k_{\mathrm{B}}T_c\ll\cutoff$. If Eq.~(\ref{eq:A1}) is also used for the calculation of $T_c^{\mathrm{3D}}$, a similar error is made and both errors can be expected to cancel in the ratio $T_c/T_c^{\mathrm{3D}}$. This cancellation works as long as the difference between $T_c$ and $T_c^{\mathrm{3D}}$ is small compared to $\cutoff$. It therefore breaks down in the limit $L\to0$, where $T_c$ diverges. The numerics shows that all subband gaps approach zero with the same slope at $T_c$, such that we have
	\begin{equation}
		\sum_qV_{1q}\frac{\Delta_q}{\Delta_1}\frac{m}{2\pi\hbar^2}\theta(\mu-E_q)
		=\frac{mV}{2\pi\hbar^2L}\left(\frac{1}{2}+N_{\mathrm{sb}}\right),
	\end{equation}
where $N_{\mathrm{sb}}$ is the number of occupied subbands. Equation~(\ref{eq:TB2}) is then readily solved to yield
	\begin{subequations}\label{eq:TBclosed}\begin{equation}
		\frac{T_c}{T_c^{\mathrm{3D}}}\approx\exp\left\{\frac{1}{\lambda}\left[1-
		\frac{(3nL^3/\pi)^{1/3}}{1/2+N_{\mathrm{sb}}}\right]\right\}.
	\end{equation}
Discontinuities occur because $N_{\mathrm{sb}}$ is a discontinuous function of $n$ and $L$. This function follows by solving Eq.~(\ref{eq:TB1}). The latter equation can be satisfied as long as the chemical potential is in the range $E_{N_{\mathrm{sb}}}<\mu<E_{N_{\mathrm{sb}}+1}$, such that one can set $\mu=E_{N_{\mathrm{sb}}}$ and solve for $N_{\mathrm{sb}}$. The result is
	\begin{align}\label{eq:Nsb}
		N_{\mathrm{sb}}&=\mathrm{floor}\left[\frac{1}{4}\left(1+C^{1/3}
			+\frac{7}{3}C^{-1/3}\right)\right]\\
			C&=3+2^5\frac{3nL^3}{\pi}+\sqrt{\left(3+2^5\frac{3nL^3}{\pi}\right)^2-\left(\frac{7}{3}\right)^3},
	\end{align}\end{subequations}
where the function $\mathrm{floor}()$ returns the largest integer smaller than its argument. Equations~(\ref{eq:TBclosed}) coincide with the black lines in Fig.~\ref{fig:fig1} up to several decimal figures. Deviations are visible only for $L\to0$ (not shown in Fig.~\ref{fig:fig1}), where $T_c$ diverges while Eqs.~(\ref{eq:TBclosed}) approach the finite value $T_c/T_c^{\mathrm{3D}}=\exp(1/\lambda)$.

\section{Weak-coupling limit of Eqs.~(\ref{eq:BCS})}
\label{app:weak}

The BCS Eqs.~(\ref{eq:BCS}) present non-analyticities that are not captured by the approximate Eqs.~(\ref{eq:TB}). As a manifestation of these non-analyticities, the three limits $\lambda\to0$, $L\to0$, and $n\to0$ do not commute. Specifically, if the limit $\lambda\to0$ is taken first, Eqs.~(\ref{eq:BCS}) reduce to Eqs.~(\ref{eq:TB}) as will be shown below. If the limit $L\to0$ is then taken in Eqs.~(\ref{eq:TB}), the resulting chemical potential approaches the bottom of the lowest subband and the resulting $T_c$ diverges. On the contrary, if the limit $L\to0$ is taken first in Eqs.~(\ref{eq:BCS}), $\mu$ approaches $E_1-\cutoff$ irrespective of the value of $\lambda$ and $T_c$ vanishes as a non-analytic function of both $L$ and $\lambda$ \cite{Valentinis-2016-2}. On the other hand, if the limit $n\to0$ is taken after the limit $\lambda\to0$, $\mu$ again approaches the bottom of the lowest subband and $T_c$ approaches a finite value, while if the limit $n\to0$ is taken first, $\mu$ approaches a value \emph{below} the lowest subband and $T_c$ approaches zero as a non-analytic function of $n$ and $\lambda$ \cite{Valentinis-2016-1}.

Here, we study the limit $\lambda\to0$ of Eqs.~(\ref{eq:BCS}) at finite $L$ and $n$. In such conditions, $\mu$ takes at $T_c=0$ the value given by Eq.~(\ref{eq:TB1}), but the relation $\mu(T_c)$ is non-analytic at $T_c=0$. A Sommerfeld-type expansion in powers of $T_c$ is therefore not possible. To study the behavior of $\mu(T_c\to0)$, we split the sum in Eq.~(\ref{eq:BCS1}) and we use the relation $\ln(1+e^x)=x+\ln(1+e^{-x})$ for the terms $q\leqslant N_{\mathrm{sb}}$:
	\begin{multline}\label{eq:BCS1bis}
		n=\frac{mk_{\mathrm{B}}T_c}{\pi\hbar^2L}\left\{\sum_{q=1}^{N_{\mathrm{sb}}}\left[
		\frac{\mu-E_{q}}{k_{\mathrm{B}}T_c}+\ln\left(1+e^{-\frac{|\mu-E_{q}|}{k_{\mathrm{B}}T_c}}\right)
		\right]\right. \\ \left.
		+\sum_{q=N_{\mathrm{sb}}+1}^{\infty}\ln\left(1+e^{-\frac{|\mu-E_{q}|}{k_{\mathrm{B}}T_c}}\right)\right\},
	\end{multline}
where we have taken into account that $E_{N_{\mathrm{sb}}}<\mu<E_{N_{\mathrm{sb}}+1}$. We define $\mu=\mu_0+\delta\mu$, where $\mu_0$ is the solution of Eq.~(\ref{eq:TB1}), which we write down for completeness:
	\begin{equation}\label{eq:mu0}
		\mu_0=\frac{\pi^2\hbar^2}{3mL^2N_{\mathrm{sb}}}\left[\frac{3nL^3}{\pi}
		+\frac{N_{\mathrm{sb}}(N_{\mathrm{sb}}+1)(2N_{\mathrm{sb}}+1)}{4}\right].
	\end{equation}
Equation~(\ref{eq:BCS1bis}) becomes
	\begin{equation}\label{eq:BCS1ter}
		n=n+\frac{mN_{\mathrm{sb}}}{\pi\hbar^2L}\delta\mu+\frac{mk_{\mathrm{B}}T_c}{\pi\hbar^2L}
		\sum_q\ln\left(1+e^{-\frac{|\mu_0+\delta\mu-E_{q}|}{k_{\mathrm{B}}T_c}}\right).
	\end{equation}
Since for all values of $q$ the exponential approaches zero for $T_c\to0$, we can use the expansion $\ln(1+x)=x$. Furthermore, except at isolated points where $\mu_0=E_q$, the correction $\delta\mu$ is negligible compared to $\mu_0-E_q$ and Eq.~(\ref{eq:BCS1ter}) can be solved to yield
	\begin{equation}
		\delta\mu\approx-\frac{k_{\mathrm{B}}T_c}{N_{\mathrm{sb}}}
		\sum_qe^{-\frac{|\mu_0-E_{q}|}{k_{\mathrm{B}}T_c}}.
	\end{equation}
We have confirmed numerically the accuracy of this expression. It shows that the deviation of the chemical potential from $\mu_0$ is exponentially small for $T_c\to0$ (or equivalently for $\lambda\to0$).

\begin{figure}[tb]
\includegraphics[width=\columnwidth]{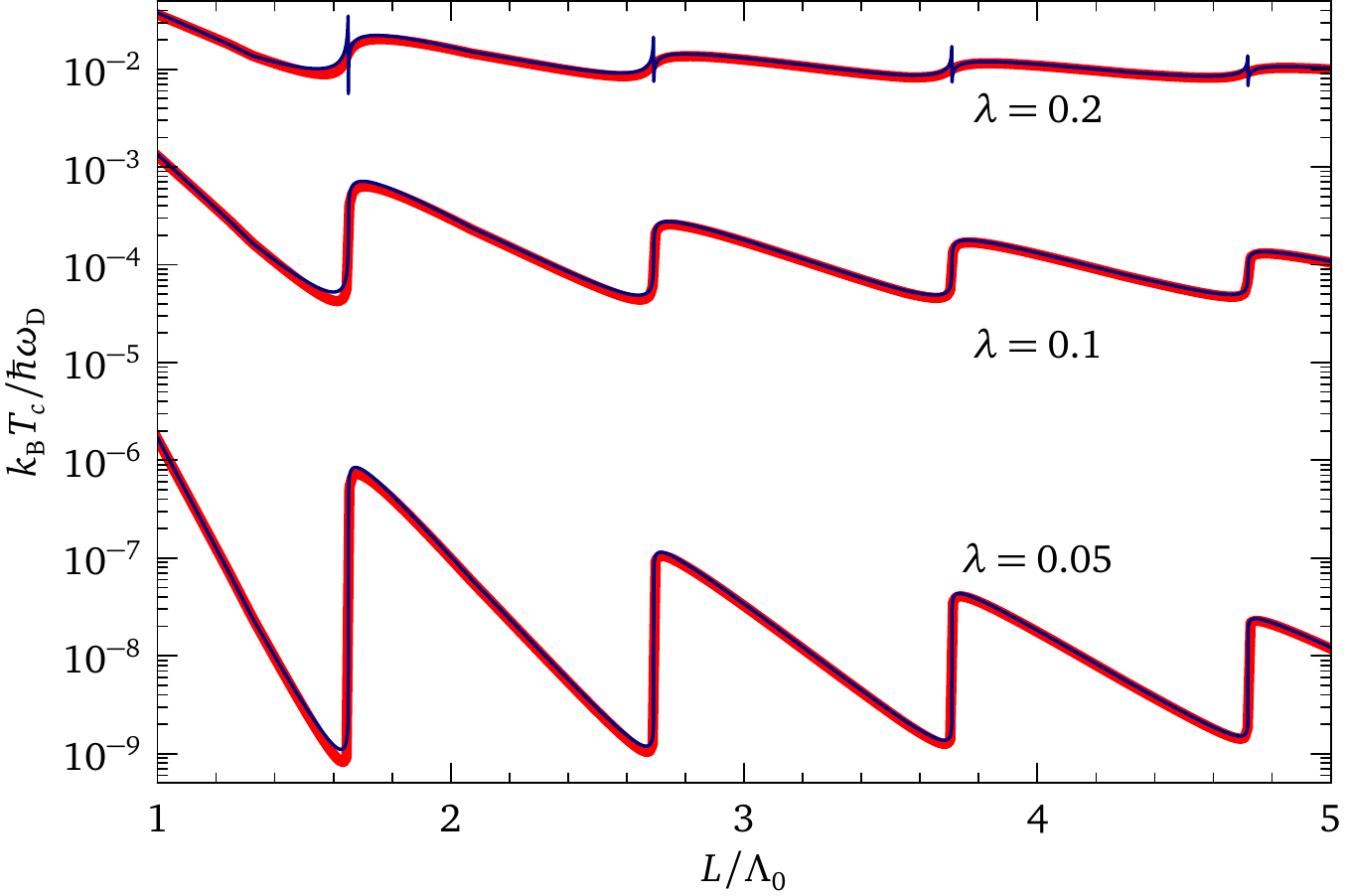}
\caption{\label{fig:fig6}
Comparison of Eq.~(\ref{eq:BCSclosed}) (thin dark-blue lines) with the numerical solution of Eqs.~(\ref{eq:BCS}) (thick red lines) for $m=m_e$, $\tilde{n}=1$ and three values of the coupling.
}
\end{figure}

We now derive a closed expression for $T_c$, which matches the solution of Eqs.~(\ref{eq:BCS}) at weak coupling and converges to Eqs.~(\ref{eq:TBclosed}) for $\lambda\to0$. If one starts from Eq.~(\ref{eq:TB2}), there are two types of corrections needed to reproduce Eq.~(\ref{eq:BCS2}). The first corrections arise from subbands such that $\mu-\cutoff<E_q<\mu$. For these subbands, Eq.~(\ref{eq:TB2}) counts the pairing of inexistent states between $\mu-\cutoff$ and $E_q$. To remove this contribution, we need the integral
	\begin{equation}\label{eq:ln1}
		-\int_{-\cutoff}^{E_q-\mu}dE\,\frac{\tanh\left(\frac{E}{2k_{\text{B}}T_c}\right)}{2E}=
		\frac{1}{2}\ln\left(\frac{|\mu-E_q|}{\cutoff}\right).
	\end{equation}
The relation Eq.~(\ref{eq:ln1}) is exact for $T_c\to0$, because $E$ is negative in the whole integration range and the hyperbolic tangent can be replaced by $-1$. The subbands that bring this correction have indices $q=N_{\mathrm{sb}}^-,\ldots,N_{\mathrm{sb}}$ with $E_{N_{\mathrm{sb}}^--1}<\mu-\cutoff<E_{N_{\mathrm{sb}}^-}$, therefore
	\begin{equation}\label{eq:Nsbm}
		N_{\mathrm{sb}}^-=1+\mathrm{floor}\left[\sqrt{\frac{2mL^2}{\pi^2\hbar^2}\left(\mu-\cutoff\right)}\right].
	\end{equation}
The corrections of the second kind arise from subbands with $\mu<E_q<\mu+\cutoff$ that are excluded from Eq.~(\ref{eq:TB2}), which therefore fails to account for the pairing of unoccupied states between $E_q$ and $\mu+\cutoff$. Adding this contribution requires the integral
	\begin{equation}\label{eq:ln2}
		+\int_{E_q-\mu}^{\cutoff}dE\,\frac{\tanh\left(\frac{E}{2k_{\text{B}}T_c}\right)}{2E}=
		-\frac{1}{2}\ln\left(\frac{|\mu-E_q|}{\cutoff}\right).
	\end{equation}
These subbands have indices $q=N_{\mathrm{sb}}+1,\ldots,N_{\mathrm{sb}}^+$ with $E_{N_{\mathrm{sb}}^+}<\mu+\cutoff<E_{N_{\mathrm{sb}}^++1}$, which implies
	\begin{equation}\label{eq:Nsbp}
		N_{\mathrm{sb}}^+=\mathrm{floor}\left[\sqrt{\frac{2mL^2}{\pi^2\hbar^2}\left(\mu+\cutoff\right)}\right].
	\end{equation}\\[-0.5em]
Proceeding as in Appendix~\ref{app:TB} and adding the corrections, we arrive at
	\begin{multline}\label{eq:BCSclosed}
		\frac{k_{\text{B}}T_c}{\cutoff}=\frac{2e^{\gamma}}{\pi}\\\times \exp\left[-\frac{
		\frac{(3nL^3/\pi)^{1/3}}{\lambda}-\frac{1}{2}\sum_{q=N_{\mathrm{sb}}^-}^{N_{\mathrm{sb}}^+}
		\mathrm{sign}(\mu-E_q)\ln\left(\frac{|\mu-E_q|}{\cutoff}\right)}{1/2+N_{\mathrm{sb}}}\right].
	\end{multline}
As the deviation of $\mu$ from $\mu_0$ is exponentially small in the weak-coupling regime, we can replace $\mu$ by $\mu_0$ in Eqs.~(\ref{eq:Nsbm}), (\ref{eq:Nsbp}), and (\ref{eq:BCSclosed}), which together with Eqs.~(\ref{eq:Nsb}) and (\ref{eq:mu0}) provide a closed expression for $T_c$. This expression compares favorably with the numerical result as seen in Fig.~\ref{fig:fig6}. Remarkably, the discontinuities contained in $N_{\mathrm{sb}}$ are precisely canceled by the correction term in Eq.~(\ref{eq:BCSclosed}) for the lowest values of $\lambda$ and the resulting $T_c(L)$ curve is smooth. At larger $\lambda$, the cancellation is imperfect and spikes appear at the thicknesses where $N_{\mathrm{sb}}$ is discontinuous. Being independent of $\lambda$, the correction term in Eq.~(\ref{eq:BCSclosed}) becomes irrelevant for $\lambda\to0$ and the expression Eq.~(\ref{eq:TBclosed}) is therefore recovered in this limit.


%

\end{document}